\documentclass[10pt,a4paper]{article}
\usepackage[nohead,width=31.5pc,left=9.9pc,top=6.5pc,bottom=10pc]{geometry}
\usepackage{amsmath}
\usepackage{amsfonts}
\usepackage{amssymb}
\usepackage{graphicx}
\author{Kavya H. Rao}
\title{Effect of double pulse irradiation on the morphology of a picosecond laser produced chromium plasma}
\begin{document}
\title{Effect of double pulse irradiation on the morphology of a picosecond laser produced chromium plasma}
\author{Kavya H. Rao$^{1,*}$, N. Smijesh$^{1,a}$, D. Chetty$^1$, I. V. Litvinyuk$^1$  and R. T. Sang$^1$}
\maketitle
\begin{flushleft}
$^1$Australian Attosecond Science Facility, Centre for Quantum Dynamics, Griffith University, Nathan, 4111, Australia.\\
$^a$Currently at the Department of Physics, Ume\r{a} University, Ume\r{a}, SE-901 87, Sweden.\\
$^*$kavya.hemantharao@griffithuni.edu.au
\end{flushleft}

\begin{abstract}
We describe the measurements to control the morphology and hence the characteristics of a picosecond laser produced chromium plasma plume upon double-pulse (DP) irradiation compared to its single-pulse (SP) counterpart. DP schemes are realized by employing two geometries wherein the inter-pulse delay ($\tau_p$) in the collinear geometry and the spatial separation ($\Delta x$) are the control parameters for schemes DP$_1$ and DP$_2$ respectively. The aspect ratio (plume length/plume width) decreases upon increasing parameters such as pressure, delay between pulses and the energy of the second pulse in DP1 scheme. Interestingly, the expansion conditions of the plume which occurs at higher pressures for SP scheme could be recreated in DP1 scheme for a lower pressure $\sim$ 10$^{-6}$ Torr. This could be potentially applied for immediate applications such as high harmonic generation and quality thin film production.
\end{abstract}

\section{Introduction}
Laser produced plasmas (LPP) have wide employability in a variety of applications \cite{phipps2007laser} such as high-order harmonic generation (HHG)\cite{ganeev2016frequency, ganeev2013high}, attosecond pulse generation\cite{Ma:16, Liu:13,attosecond}, EUV generation \cite{EUV1,EUV}, wake field acceleration \cite{wakefield,wakefield1}, material processing \cite{migliore1996}, pulsed laser deposition (PLD) \cite{Lowndes898}, nanoparticle and nanocluster generation \cite{nanoparticleapl,AmorusoPRB} etc. Despite the availability of different theories and expansion models \cite{montecarlo,blastwave,dragmodel,threeDmodel, smijesh2015dynamics}, the transient nature \cite{transient} of the plasma plume makes it difficult to predict the expansion dynamics and plume composition completely. Therefore, detailed experimental investigation of the plasma plumes by employing commonly used diagnostic techniques such as optical emission spectroscopy (OES) \cite{smijesh2016spatio,smijesh2013emission,smijesh2015dynamics}, optical time of flight (OTOF) \cite{transient}, Langmuir probe \cite{langmuir}, Thomson scattering \cite{thomson}, interferometry and shadowgraphy \cite{shadowgraphy,Noll} can be used to characterize plasmas for the applications mentioned above. In addition to this, plume imaging using an intensified charge coupled device (ICCD) \cite{smijesh2015dynamics,smijesh2014acceleration} could help to unravel the morphology and expansion of the expanding plasma plume with a better temporal resolution to about a few nanoseconds.

Attaining specific plasma parameters such as number density and plasma temperature \cite{icf,smijesh2017plasma} are crucial for most of the applications. These parameters depend on various factors \cite{hassa2016} such as the laser wavelength \cite{LPPwavelength}, pulse duration \cite{pulsewidth}, fluence \cite{LPPfluence}, spot size of irradiation \cite{LPPspotsize}, ambient gas \cite{background} as well as the material properties \cite{material}. In addition to this, irradiation schemes used for the generation of plasma also plays a major role in controlling the plume dynamics and its emission characteristics \cite{smijesh2017plasma} to a great extent. Double-pulse (DP) irradiation schemes in both orthogonal and collinear geometry has been investigated in the past to find its effect on line emission properties \cite{orthogonal, babushok2006DPreview}. However, these studies were largely carried out for nanosecond (ns) and femtosecond (fs) laser pulses \cite{mao2005DPns,sakka2012,amoruso2009plume,semerok2004ultrashort,dpf} with implications to LIBS and material processing applications. DP schemes have provided a better laser-plasma coupling depending on the nature of the preformed plasma for optimized conditions. An increase in the line emission intensity from various species \cite{DPintensity1,DPintensity2}, modest increase in the plasma temperature as well as modification of the plume morphology were reported when the DP scheme was employed with an optimal delay of $\approx$ 500-1000 ps between pulses for fs irradiations \cite{optimaldelay}. Theoretical calculations predicted an increase in the plasma temperature when the inter-pulse delay between the two pulses is $\leq$ 200 ps \cite{theorytimedelay}. Other independent investigations using DP schemes with different wavelengths\cite{DPwavelength1,DPwavelength2} indicated a drastic increase in the emission intensity and plasma temperatures (2-300 times enhancement) for an inter-pulse delay of a few microseconds to tens of microseconds (1 $\mu$s in \cite{DPwavelength1} and 25$\mu$s in \cite{DPwavelength2}). 

The current experiment aims at the investigation of picosecond laser produced chromium plasma using fast imaging techniques. Uncompressed ps pulse from the laser amplifier is widely used to generate plasmas in LPP based HHG experiments \cite{Cratto} while the compressed fs pulse drives harmonics. Attosecond pulses of $\sim$ 300 as has been previously reported \cite{Cratto} and it is possible to have phase-matched harmonic generation from Cr due to the co-existence of multiply charged species along with neutrals. Hence Cr is an important candidate for future HHG experiments and investigation of ps laser produced Cr plasma is necessary to proceed further in this direction. While experiments in the past focused on the effect of double pulse on ns and fs LPPs, inadequate information is available on the effect of DP using picosecond (ps) laser pulses. Current experiment not only investigate the characteristic expansion to bridge the gap between DP using ns pulses and fs pulses, but also reveal the use of different irradiation schemes for different applications namely, pulsed laser deposition and high harmonic generation. Collinear DP scheme (DP$_1$) is experimentally investigated by varying important parameters such as the inter-pulse delay between the first and second pulse($\tau_p$), energies in the first ($E_1$) and second ($E_2$)  pulses and the ambient/background pressure ($P_{bg}$) to investigate the distinction in the morphology of plasma plumes. In addition to this, two plasmas are generated closer to each other and the modification in the plasma plume structure is investigated by varying the separation between the two plasmas ($\Delta x$, represented as DP$_2$). The variation in velocity of fast and slow species are studied in detail along with the studies of modification in the aspect ratio (which is defined as the ratio of plume length/plume width) of the plume with respect to the variation in the above mentioned parameters.

\section{Experimental}
\begin{figure}[htbp]
\centering\includegraphics[width=11cm]{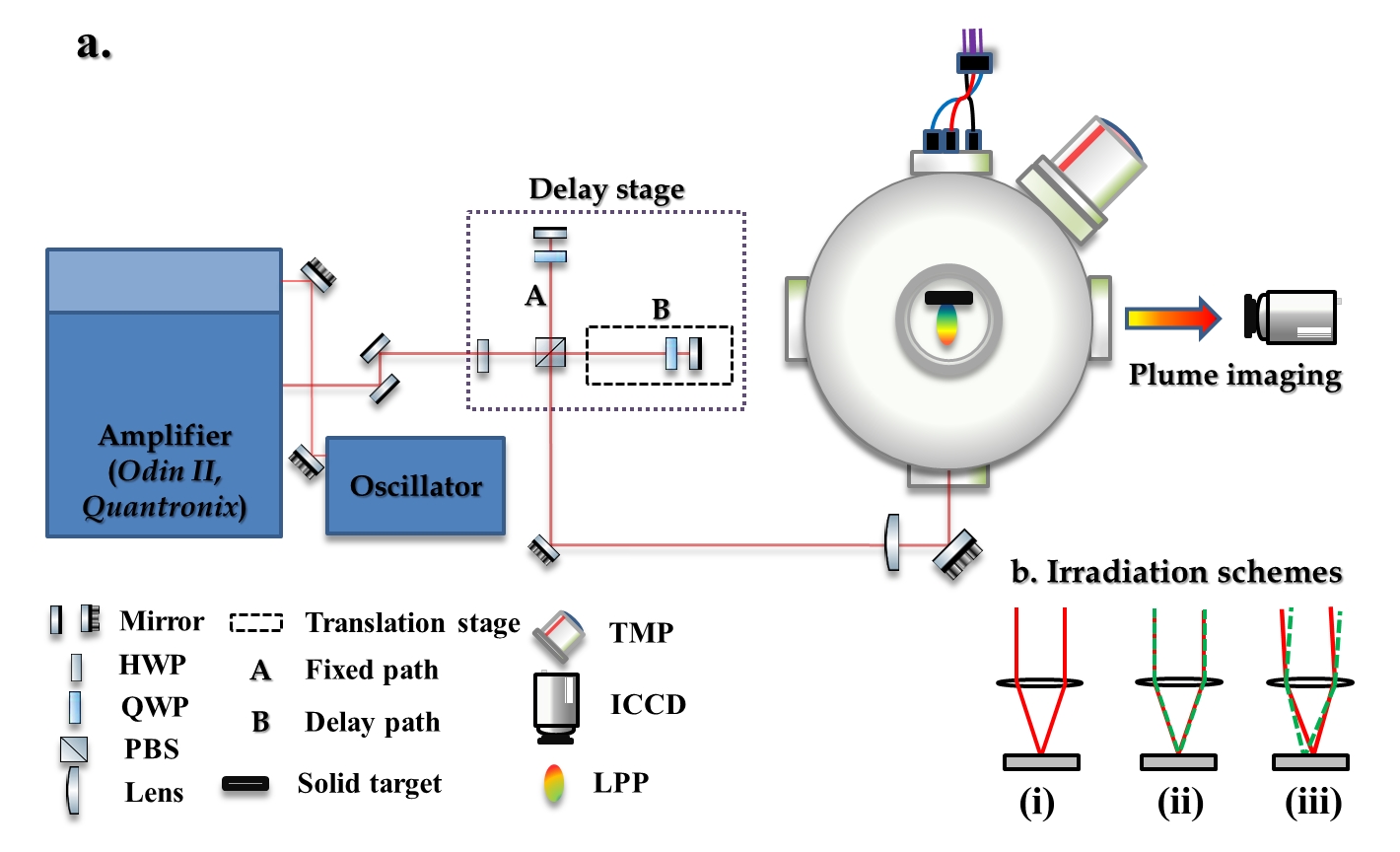}
\caption{(a.) Experimental set up for the single and double-pulse irradiation schemes for the laser plasma experiment. For double-pulse experiments, a fixed arm (labeled as A) and a delay stage (labeled as B) is set using a half wave plate (HWP) and a cube beam splitter (PBS). QWP and TMP are quarter wave plate and turbo molecular pump respectively. The whole energy is directed into either of the paths and the other path is blocked for single pulse measurements. (b.) Different irradiation schemes used; (i) Single-pulse (SP), (ii) Collinear double-pulse (DP$_1$) and (iii) DP$_2$}\label{setup}
\end{figure}

Plasma is generated by focusing a $\sim$ 60 ps laser pulses of maximum energy ($E_i$) 550 uJ from a multi-pass amplifier (\textit{Odin II, Quantronix}, operated at 1kHz) to a spot size of $\sim$ 80 $\mu$m using a 500 mm plano-convex lens onto a pure Cr (\textit{ACI Alloys Inc}, USA) target in nitrogen ambient. A fast, synchronized mechanical shutter positioned along the beam path is used to control the number of irradiations and the target is translated 200 $\mu$m in single-pulse (SP) and DP$_1$ schemes and $\geq$ 300 $\mu$m for DP$_2$ after each irradiations to avoid ablation from the pit formed by previous irradiation.  A 1024$\times$1024, Gen II ICCD (\textit{Pi:MAX 1024f, Princeton Instruments}), with a temporal resolution of $\sim$ 2 ns is used to measure the plume dynamics for different gate delays ($t_d$) with gate width ($t_w$) fixed at 10\% of the $t_d$. Double-pulse scheme is implemented by using a combination of half wave plate and a polarizing cube beam splitter as shown in figure \ref{setup}a. The path labeled as A in figure \ref{setup}a is the fixed arm and the path labeled as B is the delay arm, which allows $\tau_p$ to be varied from 0 - 1000 ps. Measurements are repeated to investigate the effect of $\tau_p$, energy in first ($E_1$)and second pulses ($E_2$) and $P_{bg}$ for DP$_1$ and the distance between two spots ($\Delta x$) in DP$_2$ and are then compared with its SP counterpart. The irradiation schemes, SP,  DP$_1$ and  DP$_2$ used in the experiments are shown in figure \ref{setup} b (i), \ref{setup} b (ii) and \ref{setup} b (iii) respectively.  A detailed description of the results on optimizing the LPP are discussed below, which has importance while employing LPP as the source medium for HHG process.

\section{Results and Discussion}
In DP, $E_i$ is divided into $E_1$ and $E_2$ (such that $E_1$ + $E_2$ = $E_i$) to ablate the Cr target from a same spot for $DP_1$ and from different spots for $DP_2$. This approach not only benefited to investigate the influence of $\tau_p$, but also find out the effect of different energy ratios between pulses on same spot so as to optimize the characteristics of the plume. Interestingly, this experiment explore the possibilities to use the pre-plasma (i.e, the plasma generated by the first pulse) as a dynamic background to confine the main plasma (i.e, the plasma generated by the second pulse) such that the main plasma expands similar to the case in SP scheme at high ambient pressures \cite{smijesh2017plasma}. On the other hand for DP$_2$, ablation from closely spaced spots on the target surface is measured for $\tau_p$ = 0 at $P_{bg}$ = 10$^{-6}$ Torr. Two plasmas generated in DP$_2$ interacts with each other on expansion leading to a completely distinct complex plume dynamics. These characteristic expansion dynamics due to the different irradiation schemes, on varying the specified parameters are discussed in detail in the following sections and with their relevances. 

\subsection{\textbf{Single pulse and double pulse irradiation schemes}}

\begin{figure}[htbp]
\centering\includegraphics[width=10cm]{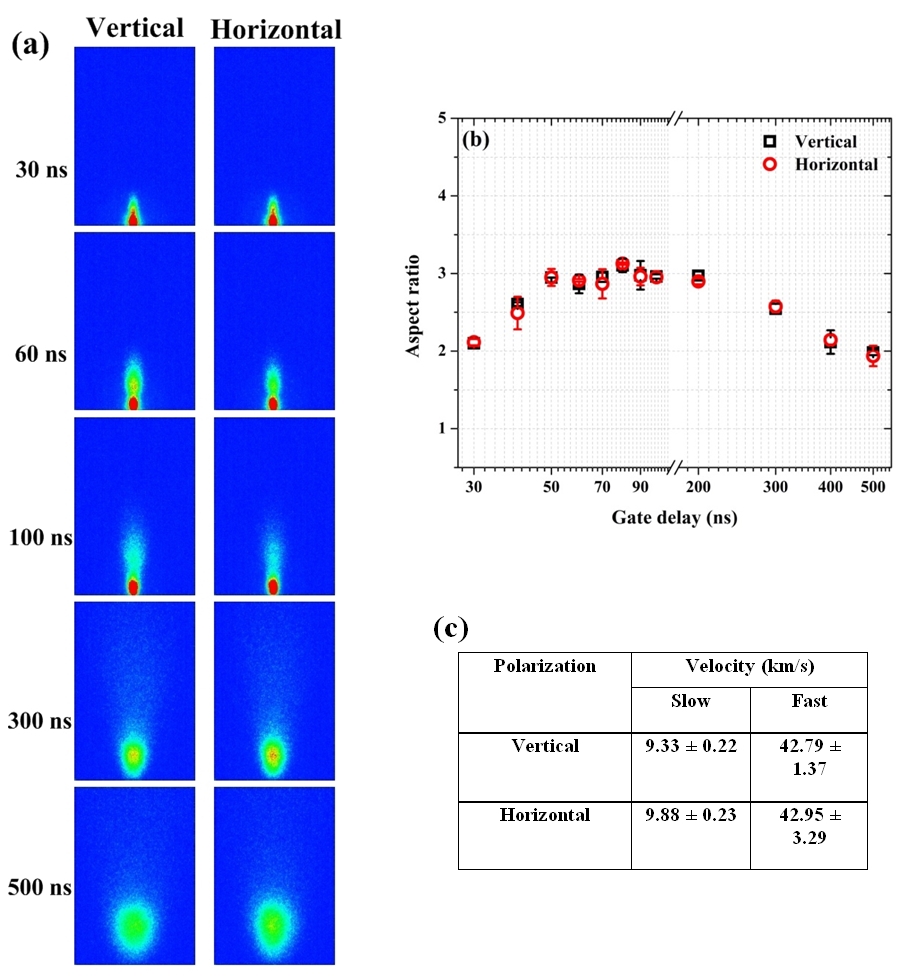}
\caption{Expansion of the plasma plume measured for vertical and horizontal polarization in SP scheme for an irradiation energy of 550 $\mu$J at a pressure of 10$^{-6}$ Torr. (a) Images of the plasma plume (2 cm $\times$ 1.5 cm ) acquired using an ICCD for various time delays ($t_d$) and integration time, $t_w$ = 10\% of the $t_d$. (b) Variation in aspect ratio (AR) with respect to $t_d$. Plume morphology is found to be unchanged for orthogonal polarizations. (c) Average velocities of the fast and slow components of the plume.}\label{polarization}
\end{figure}
Plume expansion in SP scheme with irradiation energy $\simeq$ 550 $\mu$J are recorded for various $t_d$s for orthogonal polarizations to investigate the effect of H and V polarization on the plume dynamics at a $P_{bg} \simeq$ 10$^{-6}$ Torr; the results of which are given in the figure \ref{polarization}. The plume morphology is found to be independent of  polarization (figure \ref{polarization} (a)). Plume length ($l_{LPP}$) and plume width ($w_{LPP}$) are calculated using the 1/e$^2$ times the maximum intensity in the measured plume image cross sections. These are important parameters when analyzing the plume morphology. ${l_{LPP}}/{w_{LPP}}$, the aspect ratio (AR) of the plume defines the overall plasma structure and it is investigated in detail to understand plume morphology.
Figure \ref{polarization} (b) shows the variation of AR  with respect to $t_d$ and is $\geq$ 2 illustrating that the plume expands two times along the length (normal to the target) than its width (lateral expansion) displaying a cylindrical geometry to the plume expansion. Further, the fast and slow species are found to expand with $V_f$ (velocity of fast species) $\approx$ 42 km/s and $V_s$ (velocity of slow species) $\approx$ 10 km/s and are shown in the table given in figure \ref{polarization}(c). The plume dynamics, AR of the plume, $V_f$ and $V_s$ in the plume remain unchanged upon changing the polarization. Evidently, the plume expansion dynamics is independent of the polarization of irradiation and choosing any arbitrary linear polarization doesn't influence the results; particularly when  employing DP scheme with orthogonal polarizations. 

\begin{figure}[htbp]
\centering\includegraphics[width=13cm]{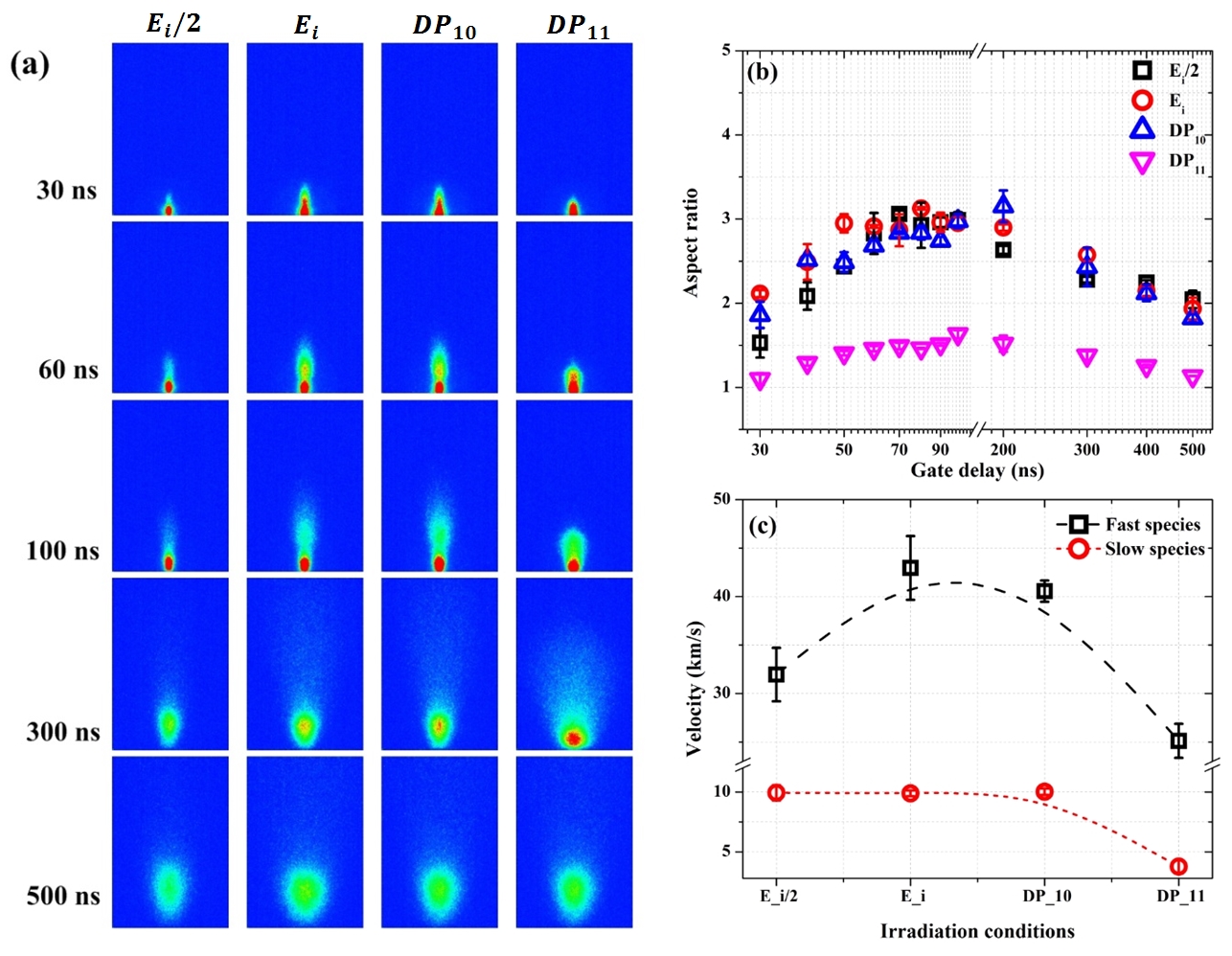}
\caption{Plume dynamics at $P_{bg}$ = 10$^{-6}$ Torr for SP and DP$_1$ schemes for a maximum irradiation energy of 550 $\mu$J. (a) 2 cm $\times$ 1.5 cm images of the plasma plume acquired using an ICCD for $t_d$ as mentioned at the left side of the figure and $t_w$ = 10\% of the $t_d$. $E_i/2$, $E_i$, $DP_{10}$ and $DP_{11}$ refers to SP at half of the maximum energy, SP at the maximum energy, DP$_1$ with $\tau_p$ = 0 ps and $\tau_p$ = 1000 ps respectively. (b) Variation in AR with $t_d$ for the different irradiation schemes presented in (a). The plume is more spherical for $DP_{11}$ when compared to all other schemes. (c) Variation in the average $V_f$ and $V_s$ in the plume as a function of the different irradiation schemes.}\label{SPnDP}
\end{figure}

Previous experiments on double-pulse scheme using fs laser pulses reported an optimum $\tau_P$ $\approx$ 500 ps -1000 ps \cite{optimaldelay} to reach a larger plasma temperature and better emission yields. Therefore, experiments are performed initially to investigate the important differences between SP and DP$_1$ schemes 
in nitrogen ambient at $P_{bg}$ = 10$^{-6}$ Torr. SP scheme is realized with energies $E_i/2$ and $E_i$, whereas DP$_1$ scheme is realized with energies $E_1$ = $E_2$ = $E_i/2$ (i.e, equal energies in both paths) with $\tau_p$ = 0 ps ($DP_{10}$) and 1000 ps ($DP_{11}$). These measurements describe that the nature of plume expansion does not change for all irradiation schemes except for $DP_{11}$ as shown in Figure \ref{SPnDP}(a). 

While plumes in SP with $E_i/2$ and $E_i$ and $DP_{10}$ are found to have similar expansion features resembling a cylindrical plasma \cite{pulsewidth} as in the case of a fs LPP plume, the $DP_{11}$ show a prominent lateral expansion resembling a ns LPP plume. To explain this further, it can be assumed that the energy of irradiation on the target by the second pulse may be $\leq$ $E_i/2$ as it passes through a preformed spatio-temporally expanded plasma (for a $\tau_P$ = 1000 ps), which might absorb/scatter a fraction of the energy depending on the cross sections of the process. This in turn let the plasma to expand laterally and can be studied by comparing the variation of AR (figure \ref{SPnDP} (b)) with respect to other irradiation schemes. While AR approaches to 1 for $DP_{11}$ illustrating a spherical morphology (which could in turn improve the plume homogeneity), the AR is $\geq$ 1.5 with a maximum of 3 at $t_d$ = 90 ns displays a larger axial expansion for all other cases. $V_f$ and $V_s$ (as shown in figure \ref{SPnDP} (c)) are calculated from the emission intensity profiles obtained from the imaging data. While the velocities of the slow species does not vary much in $E_i/2$, $E_i$ and $DP_{10}$, it decreases considerably for $DP_{11}$. Plume in $E_i/2$ reaches shorter distance when compared to $E_i$ and $DP_{10}$ as expected since plume expansion depend on irradiation energy \cite{energy-harilal} for a given LPP. $V_f$ is relatively larger and is $\approx$ 32 km/s, 42 km/s, 41 km/s and 25 km/s respectively for $E_i/2$, $E_i$, $DP_{10}$ and $DP_{11}$. The slow species move with $V_s$ $\approx$ 10 km/s for $E_i/2$, $E_i$ and $DP_{10}$, whereas it is $\approx$ 5 km/s for $DP_{11}$ and therefore the effect of $\tau_p$ on the plume dynamics for DP$_1$ needs further investigation; which will be discussed in the following sections.

\begin{figure}[htbp]
\centering\includegraphics[width=12cm]{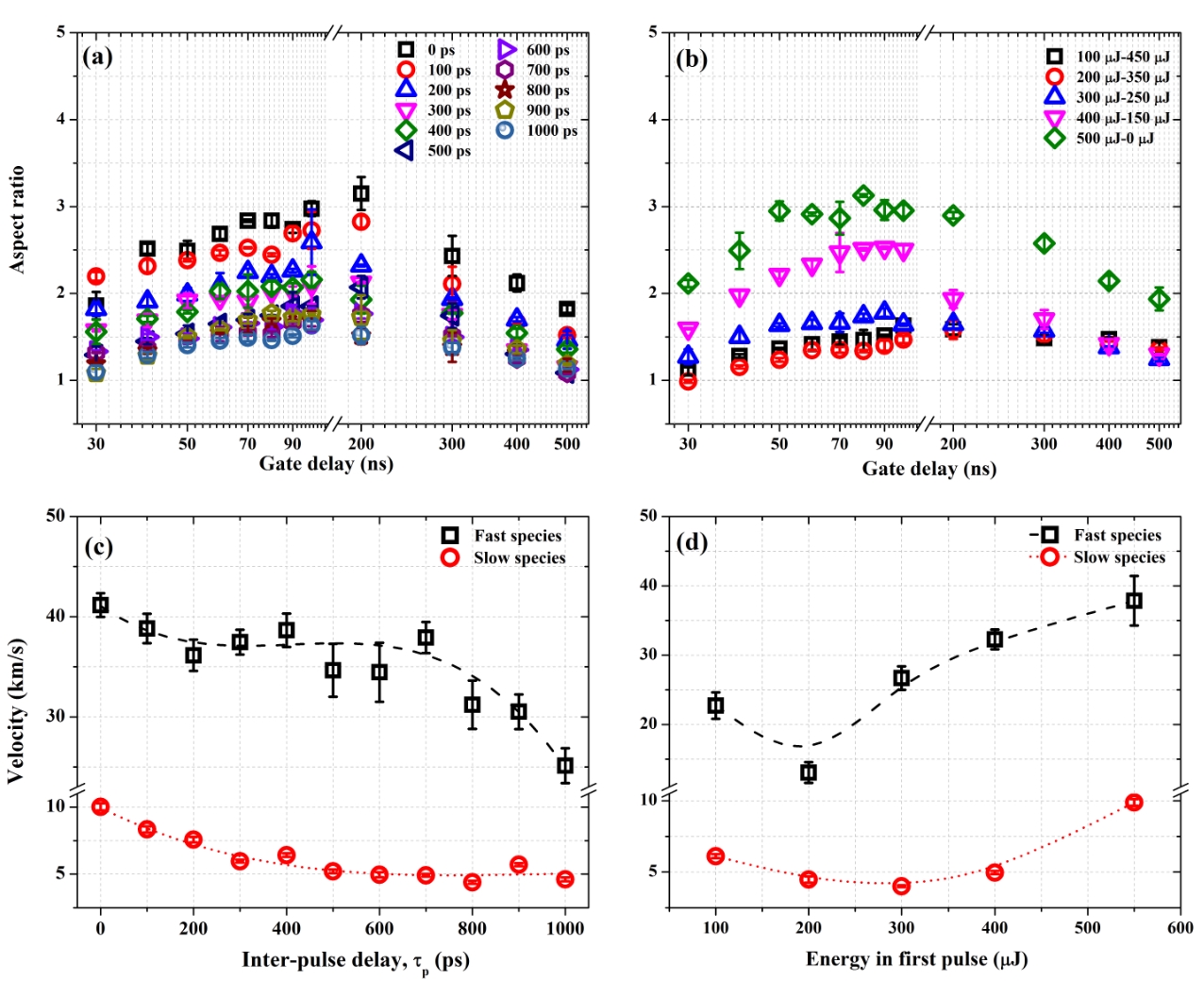}
\caption{Variation in aspect ratio of the plasma plume as a function of gate delay for (a) different inter-pulse delays, $\tau_p$ and (b) for varying energy in the first pulse, $E_1$. It can be seen that the plume is more spherical in double pulse scheme with an inter-pulse delay $\geq$ 400 ps and for cases where energy in the first pulse is less than or equal to the energy in the second pulse. Variation in the average velocity of the fast and slow components of the plume for DP$_1$ scheme with (c) different inter-pulse delays and (d) for different energy in the first pulse.}\label{delay}
\end{figure}

\subsection{\textbf{Effect of inter-pulse delay and energy in the first pulse}}

Effect of $\tau_p$ on plume hydrodynamics is investigated to find the morphological changes in the plume as well as the changes in the velocity of fast and slow species. Measurements are carried out for various $t_d$'s with $\tau_p$ varying from 0 ps to 1000 ps for $E_1$ = $E_2$ = $E_i/2$ and the results are given in figure \ref{delay} (a). The plume morphology changes from a cylindrical expansion for $\tau_p$ = 0 ps to a spherical expansion for $\tau_p$ = 1000 ps. Plume is found to exhibit a relatively low lateral expansion with AR $\approx$ 2 (for $t_d$ = 30 ns) and  $\approx$ 3 (for $t_d$ = 90 ns) for $\tau_p$ = 0 ps changes to a plume with relatively large lateral expansion with AR $\approx$ 1 (for $t_d$ = 30 ns) and $\approx$ 1.5 (for $t_d$ = 90 ns) for $\tau_p$ = 1000 ps; indicating a factor of two change in the plasma length when $\tau_p$ is varied from 0 ps to 1000ps. Interestingly, lateral expansion of the plume  is evident after $\tau_p$ = 400 ps, depicting an enhancement in the interaction of the generated second plasma produced with the pre-formed plasma, thereby increasing the possibilities for confinement. It is clear from the figure \ref{delay} (a) that the the plume length increases and reaches a maximum quickly within 100 ns after ablation and a gradual decrease for longer $t_d$s. Velocities of fast and slow components in the plume were calculated from the above mentioned measurements and is given in figure \ref{delay} (c). $V_f$ is found to decrease from $\approx$ 40 km/s to $\approx$ 25 km/s and $V_s$ decreases from $\approx$ 10 km/s to $\approx$ 5 km/s respectively on increasing $\tau_p$ from 0 ps to 1000 ps. This complements the argument on plume confinement and it could be inferred from these measurements that a longer $\tau_p$ can be chosen to have greater lateral expansion whenever required. 

Further, measurements are carried out to investigate the effect of energy $E_1$ in the $DP_1$. The energies used for irradiation was varied in such a way that the total energy of irradiation; i.e, $E_1$ + $E_2$ = $E_i$ = 550 $\mu$J. Since the previous measurements concluded that longer $\tau_p$ yields a better lateral expansion, the following set of measurements are carried out with $\tau_p$ = 1000 ps. These inter-delayed pulses generate two plasmas one after another depending on the number density ($n_c$) of the plasma formed by $E_1$ irradiation. The plume shape in these measurements are found to have considerable variation when $E_1$ is increased from 100 $\mu$J to 550 $\mu$J and the results of the variation of AR with respect to $t_d$ for different $E_1$ is given in figure \ref{delay} (b). The plume has the better confinement leading to a larger lateral expansion of the plume and hence a spherical morphology for $E_1$ = 200 $\mu$J; whereas a transition from spherical to cylindrical geometry is observed when $E_1 \geq$ 300 $\mu$J. In this case, the plasma plume generated by the first pulse expands spatio-temporally and the plasma produced by the second pulse experiences a dynamic plasma surrounding. Though the generation of the second plasma and its expansion into the pre-generated plasma depend considerably on the number density of the preformed plasma, we could record the generation of a second plasma in all our experimental conditions. For eg. when $E_1$ = 100 $\mu$J, a better ablation by the second pulse happened because of the modified heated target surface and a poor shielding of the second pulse to reach the target surface by the the pre-formed plasma due to the insufficient number density. For $E_1\leq E_2$, plasma formed by the second pulse finds a highly dynamic and transient ambient created by the plume formed from the first pulse. Complex process that involves the interaction of two plasmas may therefore result in a plume with larger radial expansion when compared to the SP counter part having similar total energy of irradiation. On the other hand, when $E_1 \geq$ 300 $\mu$J, a fraction of $E_2$ may be absorbed by the plume via inverse bremsstrahlung processes and thereby results in generating a plume with less lateral expansion when compared to the DP$_1$ case wherein $E_1 \leq E_2$. AR is closer to 1 for plasmas generated when $E_1 \leq$ 300 $\mu$J and it increases with an increase in $E_1$. Further, $V_f$ and $V_s$ were found to get reduced in DP$_1$ scheme when compared to SP with the minimum velocity of both species occurring at $E_1$ = 200 $\mu$J; which is an indication of collisional interactions/confinement happening in the plasma plume.

\subsection{Effect of ambient pressure}
Ambient pressure plays an important role in the dynamics of the plasma plume \cite{background} and hence plasma can devised for various applications by optimizing $P_{bg}$. Therefore, the experiments were repeated at different pressures for SP and $DP_1$  schemes. While SP scheme was carried out using the maximum irradiation energy available ($E_{i}$ = 550 $\mu$J), $DP_1$ scheme was carried out with $E_1$ = $E_2$ = $E_i/2$ and $\tau_p$ = 1000 ps for ambient pressures ranging from 10$^{-6}$ Torr to 10$^{1}$ Torr. 

\begin{figure}[htbp]
\centering\includegraphics[width=12cm]{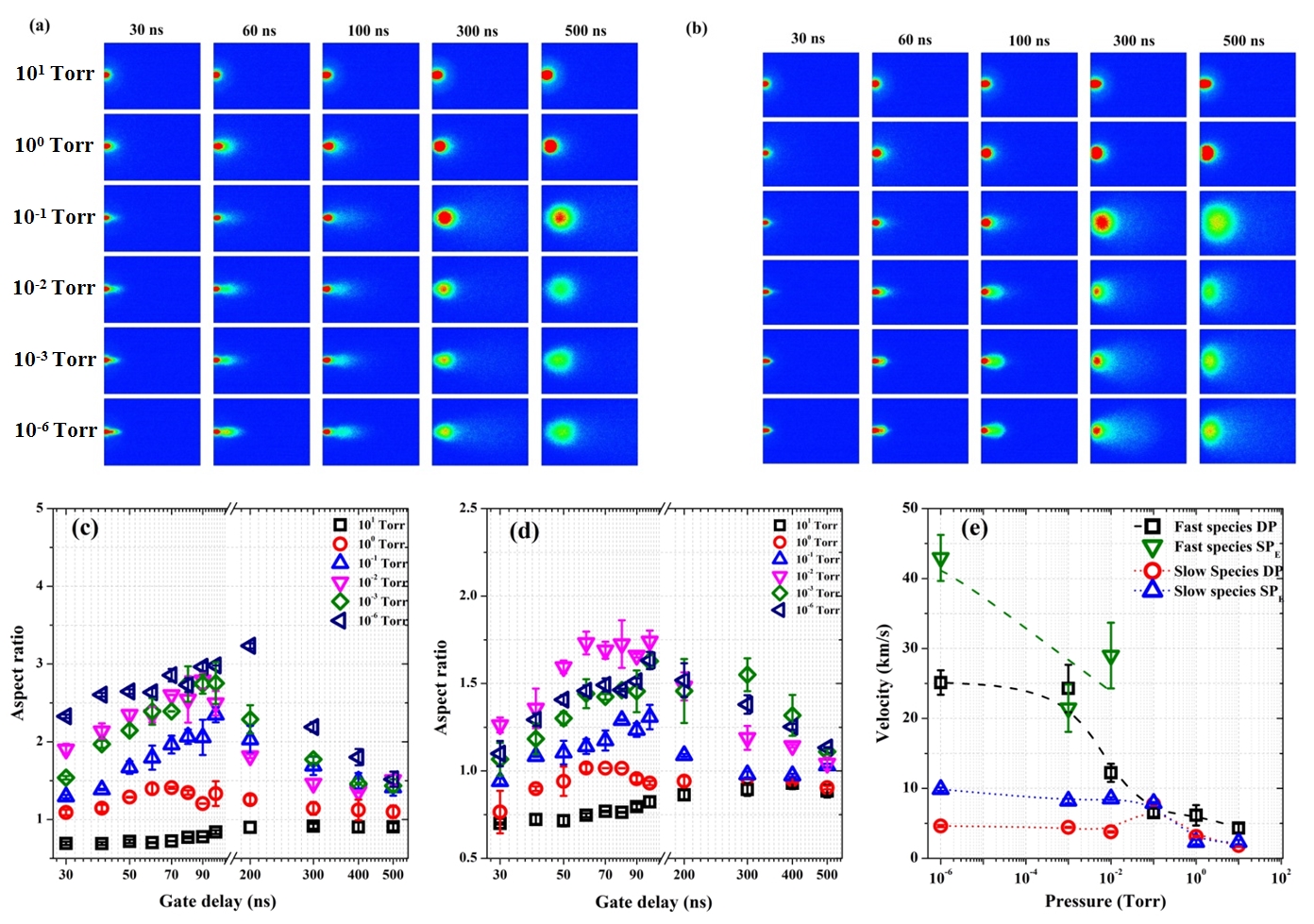}
\caption{Dynamics of the plasma plume for SP and DP irradiation schemes with respect to variation in the ambient pressure are shown here. 1.5 cm $\times$ 2 cm images of the plasma plume acquired using an ICCD for $t_d$ as mentioned in the figure legends and $t_w$ = 10\% of the $t_d$ for (a) SP scheme and (b) DP scheme. Graph showing the variation in aspect ratio with respect to $t_d$ for (c) SP scheme and (d) DP$_1$ scheme. (e) Variation in the average velocity of the fast and slow components of the plume in SP and DP$_1$ schemes.}\label{Pressure}
\end{figure}

Figure \ref{Pressure}(a) and \ref{Pressure}(b) represents the plasma plume expansion for SP and $DP_1$ schemes. Different expansion features are observed for SP and DP$_1$ scheme while $P_{bg}$ is varied. In SP, while plume expands adiabatically reaching farther distances with cylindrical plume shape (i.e. with lesser lateral expansion) at lower pressures from 10$^{-6}$ Torr to 10$^{-2}$ Torr, a sudden change to the plume structure is recorded at 10$^{-1}$ Torr with a noticeable lateral expansion. This pressure regime is considered to be really interesting and rich with many complex process has been reported in several observations \cite{smijesh2013emission}. For higher pressures, plume expansion is significantly resisted by the surrounding atmosphere leading to an enhanced lateral expansion imparting a better spherical shape for pressures 10$^0$ Torr and 10$^1$ Torr. The plume length is found to have smallest values among the current measurements, indicating a plume confinement at higher pressures for SP schemes. AR of the plume (see figure \ref{Pressure} (c)) compliments this claim since the value approaches 1 when $P_{bg}$ reaches 10$^0$ Torr from 10$^{-6}$ Torr. Interestingly at 10$^1$ Torr, plume confinement is so dominant that the AR drops to values $<$ 1 indicating a larger plume width. While slow species in the plume is present at all pressures, the fast species are visible only at pressures $\leq$ 10$^{-2}$ Torr. $V_f$ decreases with the increase of pressure, whereas for the slow species, $V_s$ remains unchanged up to 10$^{-1}$ Torr and then decreases on a further increase in pressure.

$DP_1$ experiments are repeated for pressures from 10$^{-6}$ Torr to 10$^2$ Torr displays an entirely different structure, but a similar trend for all $t_d$'s upon comparison with SP; even though the values of AR are lower. Figure \ref{Pressure} (b) shows the plume expansion dynamics and figure \ref{Pressure} (d) presents the variation in AR with respect to $t_d$ for DP$_1$ scheme. Effect of second pulse causes the plasma to expand more laterally than the axial expansion. Plume width is slightly more for DP at pressures from 10$^{-6}$ Torr to 10$^{-2}$ Torr and hence the shape of the plume found to be more spherical (with reduced AR values), when compared to the axially confined plume in SP case. Fast and slow components moves with reduced velocities (see figure \ref{Pressure} (e)) whereas AR is reduced when compared to the SP scheme. This confirms that the DP1 scheme not only confines the plume length, but also causes a lateral expansion, thereby creating a plasma that exhibit a structure similar to a ns LPP. At 10$^{-1}$ Torr pressure, it could be seen that the plume structure is more spherical at all $t_d$'s and the plume size is still bigger and brighter for $t_d \geq$ 300 ns. Though a bigger and confined plasma plume at $t_d \geq$ 300 ns is observed for SP case, the plume is much brighter and has a better spherical shape in $DP_1$ scheme. The plume is confined such that the plume width is $\geq$ the plume length for 10$^{0}$ Torr and 10$^{1}$ Torr pressures, leading to an AR $< $1. However, at later stages of expansion, it could be seen that the AR approaches 1, leading to a spherical plume at these $t_d$'s. Variation of the AR and velocities of fast and slow species in the $DP_1$ scheme are given in figure \ref{Pressure} (e).

\begin{figure}[htbp]
\centering\includegraphics[width=12cm]{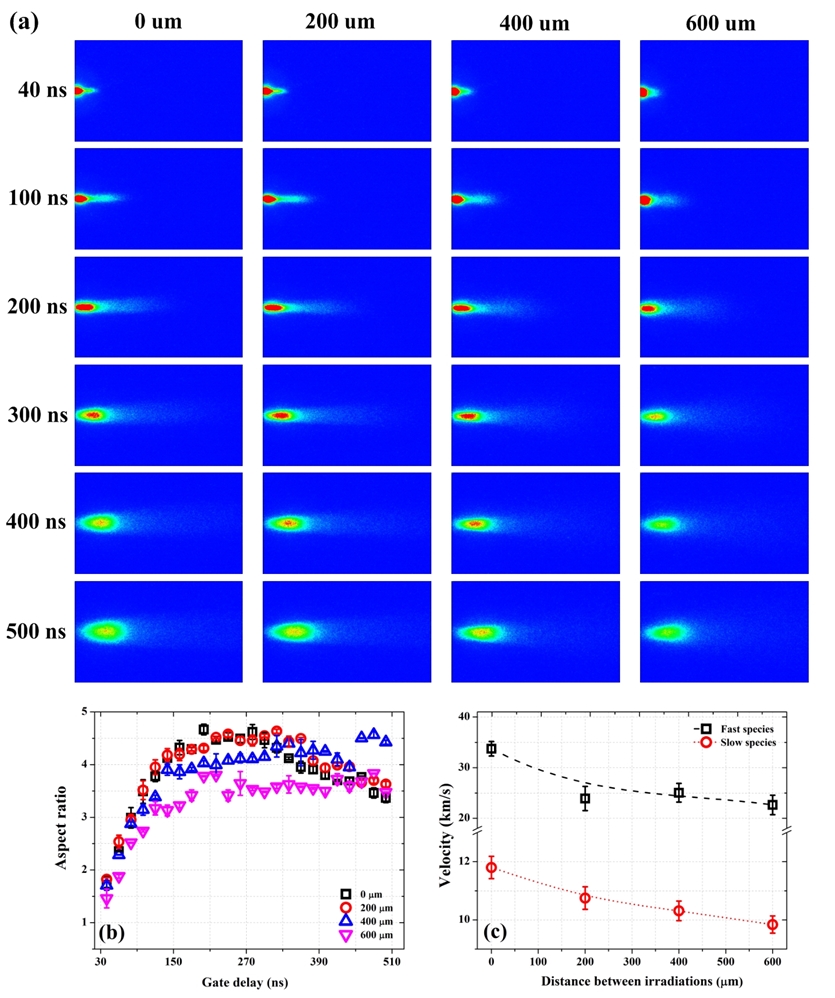}
\caption{Dynamics of the plasma plume for double pulse scheme (DP$_2$) as shown in figure \ref{setup} b (iii) at 10$^{-6}$ Torr for $E_{irr}$ = 560 $\mu$J. (a) 1.5 cm $\times$ 2 cm images of the plasma plume acquired using an ICCD for a time delay ($t_d$) as mentioned in the figure legends and integration time ($t_w$) is 10\% of the $t_d$. The distance between the laser irradiation spots ($\Delta x$) is given on the top of each column. (b) Variation in aspect ratio with respect to $t_d$ for different $\Delta x$. (c) Variation in the average velocity of the fast and slow components of the plume with respect to $\Delta x$.}\label{DP2}
\end{figure}

\subsection{Double-pulse at different spatial points}

\begin{figure}[htbp]
\centering\includegraphics[width=10cm]{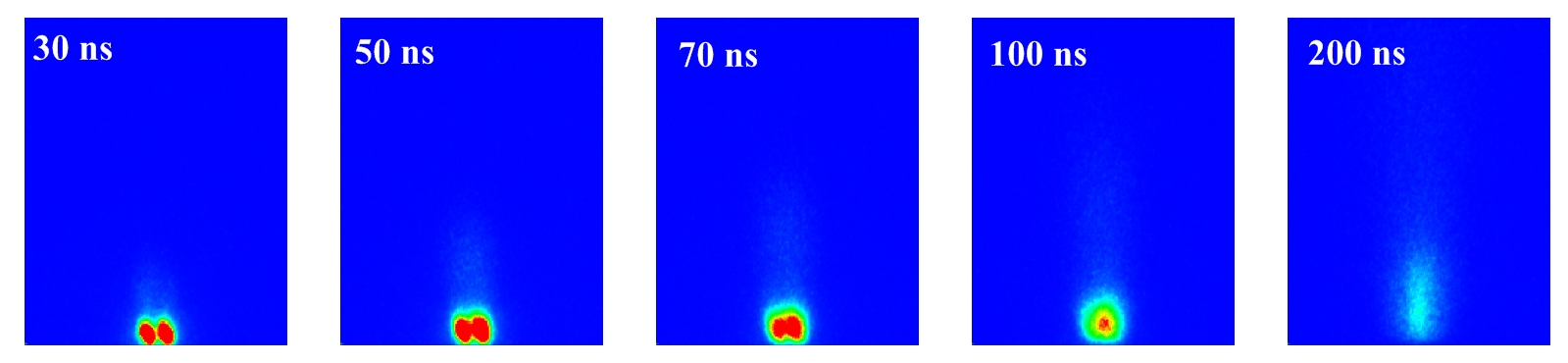}
\caption{Time-gated images of the plasma plume evolution captured for a $t_d$ as mentioned in each picture and $t_w$ = 10\% of $t_d$. The images are captured for $E_1$ = $E_2$ = $E_i/2$, $\tau_p$ = 0 ps, $P_{bg}$ = 10$^{-6}$ and $\Delta x$ = 600 $\mu$m. Dimension of image is 1 cm $\times$ 0.8 cm.}\label{DP2-600um}
\end{figure}

Apart from the DP$_1$ scheme wherein parameters like $\tau_p$, variation in $E_1$ and $E_2$ and $P_{bg}$ were optimized to tailor the plume morphology, another double pulse scheme, $DP_2$ was employed to study the variation in the plume morphology. In this scheme, plasmas are generated on the target using two identical pulses of energies $E_1$ = $E_2$ = $E_i/2$ with $\tau_p$ = 0 ps for a $P_{bg}$ = 10$^{-6}$ Torr, such that both pulses reach the target surface at the same time and simultaneously creates two closely separated plasmas with a separation = $\Delta x$. Figure \ref{DP2} (a) shows the expansion features of the plasma plumes when $\Delta x$ is varied from 0 $\mu$m to 600 $\mu$m. The plasma is found to be highly directional and cylindrical in shape for all cases and directionality increases on increasing $\Delta x$. From careful detailed analysis it is found that the plume width and plume length increases at constant rate causing the AR to remain constant soon after it reaches a maximum value and it persists for longer $t_d$s. It is also found that when $\Delta x$ is increased from 0 $\mu$m to 600 $\mu$m, the maximum value of the AR  decreases as shown in Figure \ref{DP2} (b). While the $V_f$ shows an intermediate value between SP and $DP_1$ schemes, $V_s$ displays larger velocities in comparison with the other two schemes. More specifically, $V_f$ decreases when $\Delta x$ is increased from 0 $\mu$m to 200 $\mu$m and thereafter remains almost constant for all other $\Delta x$'s. $V_s$, however decreases by increasing $\Delta x$ and it is found to have longer life time in the plume, thus forming major component of the plume in all cases. To understand the process further, the plume interaction and formation of the directed plume in $DP_2$ case, the expansion of plasma plume when $\Delta x$ = 600 $\mu$m is analyzed in detail and is given in figure \ref{DP2-600um}. 

When the two pulses hit the target surface, two distinct plumes are formed as in the figure \ref{DP2-600um}, (with legend 30 ns) and they propagate rather independently for a given time and space. On expansion these two plasmas can interact along their edges as in figure \ref{DP2-600um},for 50 ns and 70 ns, leading to the creation of a stagnation layer between the two plasmas which initiate the process of collisional interaction between them\cite{stagnation_layer}. As a result of the interaction between fast and slow components at different times, the interaction becomes more complex leading plume to move together, thereby having a longer persistence of slow components. Due to the interaction along the edges of the plasma for $t_d \geq$ 50 ns, the AR has a larger value and these interactions imparts a highly directional and cylindrical nature for the plume expansion. These kind of plasmas could be suitably used for the production of quality thin films as plume dynamics displays a better cylindrical expanding plume shape, that attains a constant plume length after $\approx$ 200 ns with better homogeneity in its cross section perpendicular to the axial expansion direction.

\section{Conclusion}
Plasmas generated by laser pulses of 60 ps duration at 800 nm, are studied under different irradiation schemes. Single-pulse (SP) and double-pulse (DP) schemes are used to understand their effects on the plasma plume morphology, which in-turn would help devising them for applications as in high-harmonic generation which is considered here. Double-pulse scheme has been carried out in two different ways: back to back ( or collinear) irradiation of the pulses on the same spatial point ($DP_1$) and irradiation of the two pulses at slightly different spatial points ($DP_2$). Plume morphology in all these cases are different and are found to differ significantly by varying the inter-pulse delay ($\tau_p$), energy in the first and second pulse ($E_1$ and $E_2$) and background pressure ($P_{bg}$) for DP$_1$ and with respect to the spatial separation ($\Delta x$) for DP$_2$. DP$_1$ helps in obtaining a more spherical plasma, with aspect ratio close to 1 when compared to SP and DP$_2$ scheme that generates a cylindrical plasma where aspect ratio is greater than 1 for $\tau_p \geq$ 400 ps. Also, a spherical plume with aspect ratio greater than 1 is observed When the energy in the first path is less than or equal to the energy in the second path. For SP schemes wherein the background pressures are varied, plume confinement is observed on an increase in the background gas with an aspect ratio closer to 1 for higher pressures. Furthermore, the effect of double pulse with an increase in background yields to a more spherical expansion when compared to their respective SP counterparts. Therefore, the plume expansion conditions as in the SP case with higher $P_{bg}$ can be reproduced in $DP_1$ scheme at lower pressures by wisely choosing $\tau_p$, $E_1$ and $E_2$. This would help in improving the temperature and number densities of the plasmas at relatively lower pressures and also ensure a more homogeneous plasma that would support phase matching/ quasi-phase matching conditions for efficient high-order harmonic generation from plasmas. Whereas plume morphology is entirely different in DP$_2$ when compared to that of SP and DP$_1$. A more directed and cylindrical plume with aspect ratio $>$ 1 is always observed, which could be suitably utilized for producing quality thin films by laser ablation/plasma-assisted thin film deposition. 

\section*{Acknowledgements}
This project is funded by the Australian Research Council Linkage project grant No. LP140100813. Kavya H. Rao was supported through an ``Australian Government Research Training Program Scholarship" and N. Smijesh was supported by the Griffith University Postdoctoral Fellowship Scheme.

\end{document}